\documentclass[twoside]{ilcws10}
\usepackage[latin1]{inputenc}
\usepackage[dvips]{graphicx,epsfig,color}
\usepackage{graphicx}
\usepackage{wrapfig,rotating}
\usepackage{amssymb,amsmath,array}
\usepackage{hyperref}

\begin{document}
\title{What can we learn at ATF2 concerning ILC backgrounds ?} 
\author{Hayg Guler$^1$, Marc Verderi$^1$ 
\vspace{.3cm}\\
1- LLR-Ecole Polytechnique - ATF2 \\
Route de Saclay, Palaiseau - France
\vspace{.1cm}\\
}

\maketitle

\begin{abstract}
  The ATF2 project aims at demonstrating the strong vertical electron
  beam focusing capability, down to the few tens of nanometers level,
  of a down scale prototype of the final focus system of the next
  generation of e+e- machines. ATF2 offers opportunities to check in a
  real accelerator environment case for the performances of the beam
  transport and background generation code, used in the simulation of
  the future e+e- machines, BDSIM \cite{BDSIM}, and for the performances of its
  underneath particle-matter simulation code, Geant4.

\end{abstract}

\section{Background sources : From ILC to ATF2}
As an electron-positron collider, the ILC may provide a very clean
experimental environment compared to hadron colliders, but it is
certainly not background-free. The rates for events from high-energy
electron-positron interactions are low: at the nominal luminosity of
$2.10^{34} cm^{-2}s^{-1}$, there will be less than one hard
electroweak interaction per second at 500 GeV, even for processes that
are not in the main focus of physical analyses. Consequently, the most
important source of unwanted interactions are machine induced
backgrounds. This term denotes all particles that are produced due to
the operation of the accelerator itself and due to collective effects
from the collision of the particle bunches as a whole. Various
sources of background can be studied at ATF2, in order to test and
improve the simulation codes and methods, which can be useful to
understand and evaluate for next Linear Collider some sources of
backgrounds.  
\begin{itemize}
\item Machine produced background before IP
\item Beam beam background at IP
\item Spent beam background 
\end{itemize}

\subsection{Machine-related backgrounds}
The background level due to the machine needs careful evaluation and
development of means to reduce it. It can to a large extend be
influenced by the design of the beam delivery system. 
\subsubsection{Beam tails (Halo) from linac}
Particle fluxes generated in the interactions of a beam halo with the
collimators in the ILC Beam Delivery System (BDS) can exceed tolerable
levels for the collider detectors and create hostile radiation
environment in the interaction region (IR hall). At ATF2, the beam
tail (halo) generated background is an important source of background
for beam diagnostic systems, and one way to reduce it, is to use
dedicated collimators. The beam halo has also been measured in order
to improve the simulations of the background since the beam profile
enters as an input parameter for the simulation.
\begin{figure}[ht]
\begin{minipage}[b]{0.5\linewidth}
\centering
\includegraphics[scale=0.55]{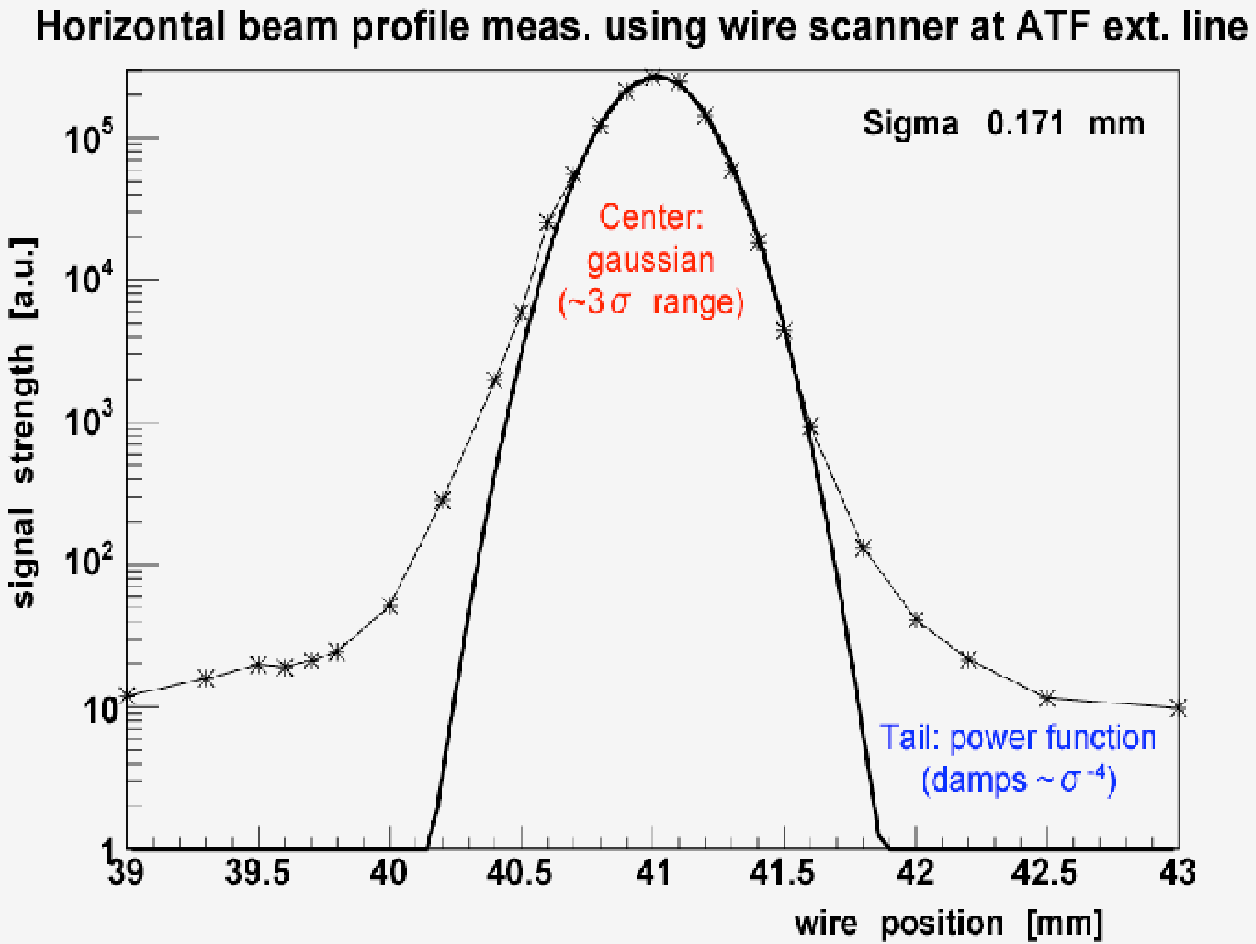}
\caption{Beam Halo measured at ATF using wire scanners.}
\label{fig:Halo_meas}
\end{minipage}
\hspace{0.5cm}
\begin{minipage}[b]{0.5\linewidth}
\centering
\includegraphics[scale=0.55]{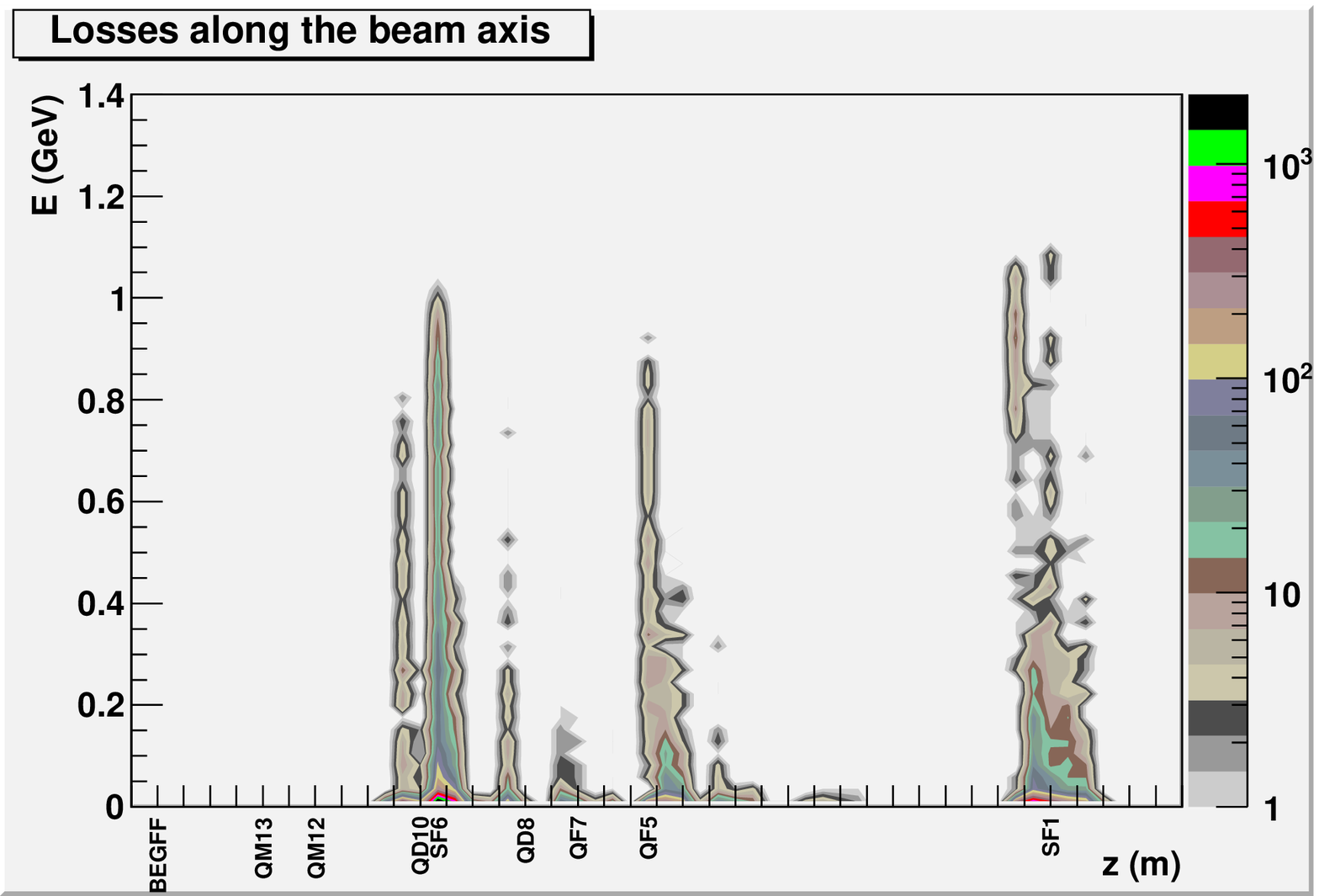}
\caption{Beam losses along ATF2 line: BDSIM simulation using as imput a Gaussian beam distribution
  for the core with 10 $\sigma$ Gaussian tails for the halo.}
\label{fig:Losses_vs_Z}
\end{minipage}
\end{figure}
\subsubsection{Synchrotron radiations}
This is major background at FFTB experiment, but at ATF2, critical
energy at bending magnet (assuming 1 Tesla) is about 1 keV. As a
result, synchrotron photons can be easily stopped by beam pipe. 
\subsubsection{Beam-gas scattering}
The four possible mechanisms of beam scattering \cite{beam_gas} are:
the scattering from blackbody thermal photons, inelastic scattering
from residual gas molecules (beam-gas Bremsstrahlung), elastic
beam-gas scatterings (relativistic Coulomb scattering) and the
scattering off atomic electrons. The background generated by the
beam-gas scattering is negligible at ATF2 since the beam line is
relatively short (less than 100m), but can contribute to the beam halo
due to large angle scattering. 
\subsection{Beam-Beam background @ IP}
A major contribution to machine-induced backgrounds are
electron-positron pairs that are created in the scattering of
beamstrahlung photons. Since this background needs two beams, it
cannot be accessed and studied at ATF2.

\subsection{Neutron background}

Neutrons will be produced wherever particles are lost, but
predominantly by the high-power beam in the beam dump. The hottest
regions of the whole accelerator complex clearly are the beam dumps,
which have to absorb a total power of approximately 10 MW each. Note
that the 500 GeV machine beam power is 11MW and the 1 TeV machine beam
power is 18MW \cite{DUMP}. In the case of water dumps, the incident particles will
produce neutrons in amounts of the order of $10^{12}$ particles per
bunch crossing. Earlier studies \cite{neutron1} showed that a small
amount of these neutrons (around $10^7$ particles) can escape from the
dump and move back into the beam tunnel, but again only a tiny
fraction (around $10^{4}$ particles) can reach the detector. \\
The ATF2 1.3 GeV beam ends up in an iron made dump, generating low
energetic neutron and electromagnetic background. Measuring this
background is a good opportunity to test for the performances of
background generation codes used to predict ILC background levels.

\section{Background simulation and measurement @ ATF2}
ATF2 (Accelerator Test Facility 2)\cite{atf2} is a final focus test
bench for ILC. It has 2 major goals, which are achievement and
maintenance of 37 nm beam size by ILC-like beam optics and
stabilization of beam position to nanometer level.\\
At ATF2, multiple source of background enters quasi-simultaneously the
beam tunnel making difficult to use the beam diagnostic systems. On
the other hand, that multiple background represent a good opportunity
to test in accelerator environment simulation tools that are used to
simulate the next e+e- background.

\subsection{Measurement apparatus}
To disentangle the various sources of background, a simple and
flexible apparatus system has been designed and built : it is made of
a set of readout modules that can either be grouped in a
longitudinally segmented mini-calorimeter, with the possibility to
insert radiator medium, or used individually for synchronized
multi-point measurements. Each module is made of one crystal (plastic
scintillator or pure CsI) and a fast photomultiplier tube. Both
crystals are sensitive to electromagnetic background and neutron
signals as well, with difference concerning the neutron background.
Neutrons do not produce ionization directly in scintillation crystals,
but can be detected through their interaction with the nuclei of a
suitable element. Fast Neutrons can interact with materials that
contain a large concentration of hydrogen atoms (protons), for example
organic materials, by means of elastic scattering in which case the
energy of the neutron is (partially) transfered to the protons which
on their turn can produce scintillation light. Using the above
principle, fast neutrons can be detected in any organic (plastic)
scintillator. From the pulse shapes measured at ATF2, it comes out
that plastic scintillators are more sensitive to fast neutrons and CsI
crystals can see large neutron energy spectrum and are useful to
measure slow (thermal) neutrons.\\
In order to discriminate neutrons from electromagnetic background, the
particles time of flight is used since the neutrons are in general
delayed from direct electromagnetic background.


\begin{figure}[ht]
\begin{minipage}[b]{0.5\linewidth}
  \centering
  \includegraphics[scale=0.55]{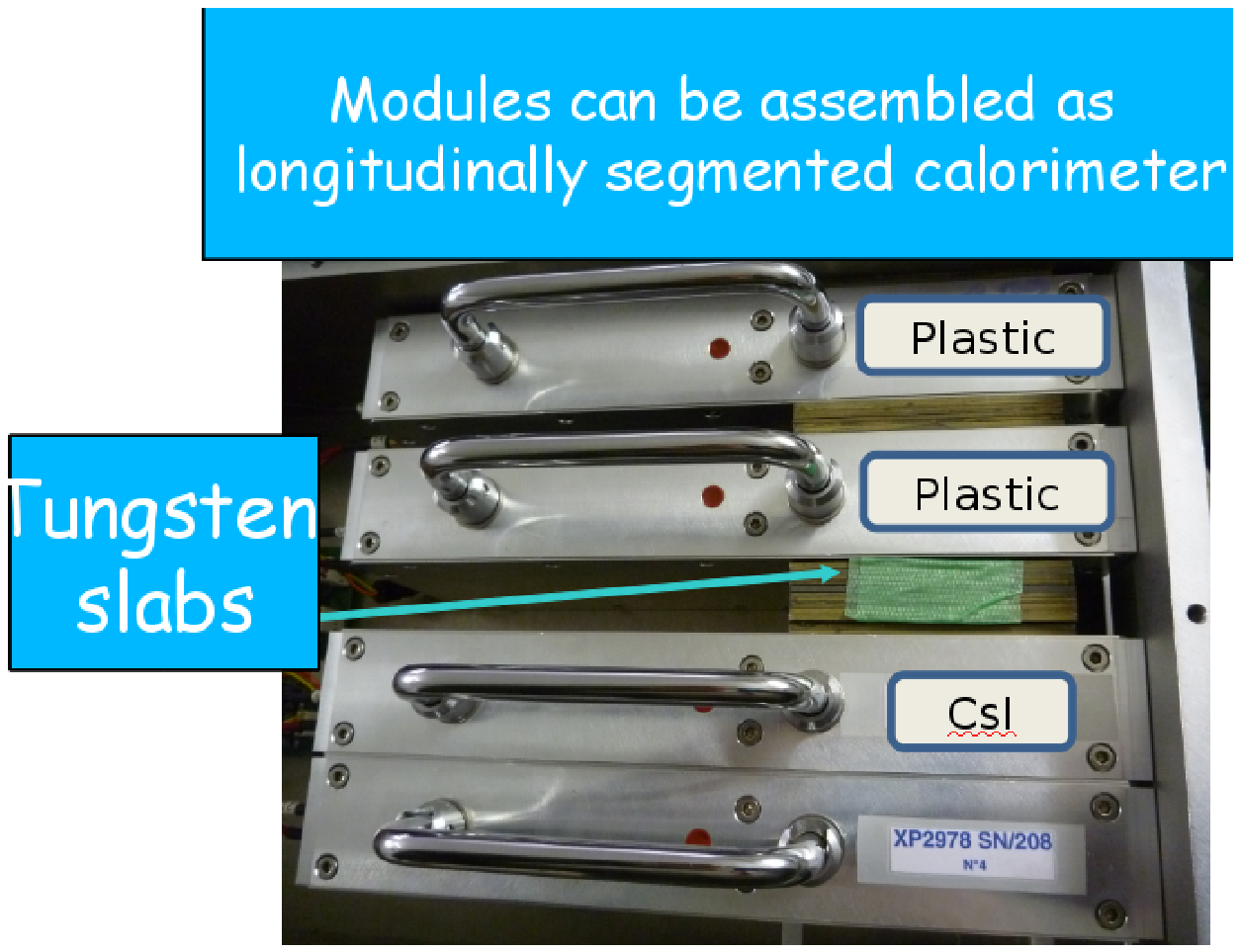}
  \caption{Picture of the built modules in a mini-longitudinal
    calorimeter configuration. Tungsten slabs are used as radiator
    medium. }
  \label{fig:1}
\end{minipage}
\hspace{0.5cm}
\begin{minipage}[b]{0.5\linewidth}
  \centering
  \includegraphics[scale=0.55]{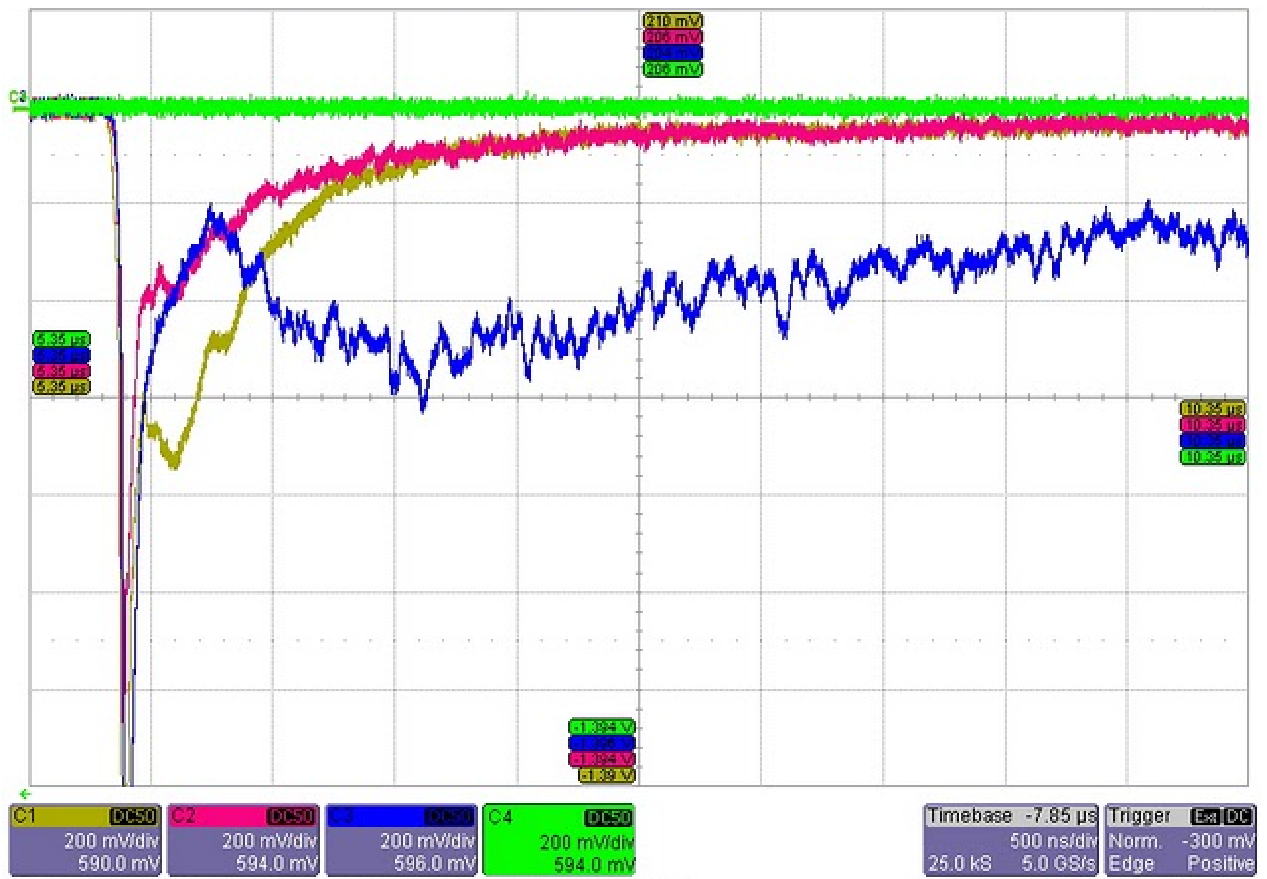}
  \caption{First signals at ATf2 : pink and yellow signals are coming
    from the modules containing plastic scintillators and the blue
    curve corresponds to a module containing pure CsI crystal.}
  \label{fig:2}
\end{minipage}
\end{figure}

\subsection{Simulation tools}
The simulations are done with BDSIM and stand alone GEANT4\cite{geant4}
codes for specific studies. BDSIM is a code that combines
accelerator-style particle tracking with traditional Geant-style
tracking based on Runge-Kutta techniques. This approach means that
particle beams can be tracked efficiently when inside the beampipe,
while also enabling full Geant4 processes when beam-particles interact
with beamline apertures. Tracking of the resulting secondary particles
is automatic. 
\begin{figure}[!h]
  \centering
  \includegraphics[scale=0.6]{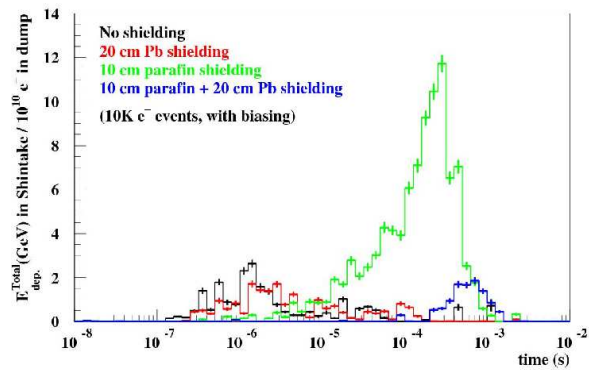}
  \caption{Simulation of the neutron background on the Shintake photon
    detector. Specific event biasing methods have been used in order
    to have enough statistics in the detector, and understand the
    impact of dedicated shielding.}
  \label{fig:1}
\end{figure}
A standard simulation efficiency is relatively poor as soon as one
need enough statistics on a detector placed off beamline axis. Indeed,
the statistic is mainly focused around the beam core inside the
beampipe. One way to enhance the counts in a given off axis detector
is to generate enough particles by consuming computing time to
compensate the low cross sections for any unlikely situations. Geant4
provides event biasing techniques which may be used to save computing
time. Such techniques have been used in order to simulate the neutron
background in the gamma detector used at ATF2 in the nanometer beam
size measurement, the Shintake monitor\cite{Shintake}.

\section{Summary and outlook}
Background measurement program has been started at ATF2 in order to
test in real accelerator environment simulation tools that are used
to simulate the next ILC background levels. A dedicated apparatus,
specific to small ATF2 final focus area, specific to ATF2 multiple
background, has been studied and built. First background measurements
have been done in 2009 and specific simulation have been done in order
to understand the measurements.\\
Next years will be dedicated to make other dedicated measurements and
put constraints on the simulation tools BDSIM and GEANT4.

\begin{footnotesize}
  
  
\end{footnotesize}


\end{document}